\documentclass[12pt]{iopart}
\usepackage{graphicx}%
\usepackage{dcolumn}
\usepackage{iopams}   
\usepackage{color}
\usepackage{bm}
\usepackage{longtable}
\usepackage{subfig}
\begin{document}
\newcommand{\text}{\mathrm}

\title[Protein dielectrics]{On the Theory of Dielectric Spectroscopy of Protein Solutions }
\author{Dmitry V.\ Matyushov}
\address{Center for Biological Physics, Arizona State University, 
PO Box 871504, Tempe, AZ 85287-1504
}
\ead{dmitrym@asu.edu}

\begin{abstract}
  We present a theory of the dielectric response of a solution
  containing large solutes, of a nanometer size, in a molecular
  solvent. It combines the molecular dipole moment of the solute with
  the polarization of a large subensemble of solvent molecules at the
  solute-solvent interface. The goal of the theory is two-fold: (i) to
  formulate the problem of the dielectric response avoiding the
  reliance on the cavity-field concepts of dielectric theories and
  (ii) to separate the non-additive polarization of the interface,
  jointly produced by the external field of the laboratory experiment
  and the solute, from specific solute-solvent interactions
  contributing to the dielectric signal. The theory is applied to
  experimentally reported frequency-dependent dielectric spectra of
  lysozyme in solution. The analysis of the data in the broad range of
  frequencies up to 700 GHz shows that the cavity field
  susceptibility, critical for the theory formulation, is consistent
  with the prediction of Maxwell's electrostatics in the frequency
  range of 10--200 GHz, but deviates from it outside this range. In
  particular, it becomes much smaller then the Maxwell result and
  shifts to negative values at small frequencies. The latter
  observation implies a dia-electric response, or negative
  dielectrophoresis, of hydrated lysozyme. It also implies that the
  effective protein dipole recorded by dielectric spectroscopy is much
  smaller than the value calculated from protein's charge
  distribution. We suggest an empirical equation that describes both
  the increment of the static dielectric constat and the decrement of
  the Debye water peak with increasing protein concentration. It gives
  fair agreement with broad-band dispersion and loss spectra of
  protein solutions, but misses the $\delta$-dispersion region.
\end{abstract}

\noindent
\submitto{\JPCM}
\maketitle

\section{Introduction}
Dielectric spectroscopy is a linear response technique, monitoring the
dynamics of the dipole moment of a macroscopic sample of a polarizable
material \cite{Bottcher:78,Takashima:89}. While it is highly sensitive
and provides wealth of information about the dynamics of polarization
modes active in a medium, the interpretation and the assignment of the
observed relaxation processes often require theoretical approaches.

The standard theoretical tool to study mixtures is the Maxwell-Wagner
theory \cite{Takashima:89,Scaife:98} and its modifications, also in
terms of effective-medium approaches \cite{Choi:99}. All these theories
assume that macroscopic dielectric constants can be assigned to all
components of the mixture. This often becomes a significant
oversimplification when highly heterogeneous solutes of nanometer
dimension, such as hydrated proteins, are
involved \cite{Smith:93,Pitera:01}. The description of the polar
response in terms of the molecular charge distribution is more accurate
for these solutes \cite{King:91}. Given the length-scale of the
external field variation in a typical dielectric or light-absorption
experiment, the overall charge and the dipole moment are the two main
multipoles to consider \cite{South:72}.

Once a dipole is assigned to a protein, one might assume that standard
models of dipolar liquids \cite{SPH:81,Madden:84}, involving the
statistics and dynamics of molecular dipoles, can be directly extended
to study protein solutions. One has, however, to recognize that
proteins, and other solutes of similar dimension, possess an extended
interface with a molecular solvent, such as water, which is absent in
the case of mixtures composed of molecules of comparable size. The
interface of a hydrated protein involves a large number, $\sim
300-500$, water molecules only in the first hydration layer. Given
that the perturbation of the water polarization propagates at least
into the second hydration layer \cite{DMcpl:11,DMjcp1:12}, the actual
size of the protein-water interface is significantly larger.

These new physical realities pose the requirement to develop new
theoretical approaches to describe the polar response of protein
solutions. The key question for this development is how to extend the
classical theories of polar response of molecular dipoles into the
realm of large solutes with an extended interfaces.  The key parameter
for the development of dielectric theories is the Onsager cavity (or
directing) field \cite{Onsager:36} producing the torque on the
molecular dipole when the macroscopic sample is placed in an external
electric field \cite{Boettcher:73}.  The standard result of the
classical theories is that the field of external charges $E_0$ is
screened by the polarization of the interface to the cavity field
\cite{Boettcher:73,Jackson:99}
\begin{equation}
  \label{eq:17}
  E_c = \frac{3}{2\epsilon_s+1} E_0 ,
\end{equation}
where $\epsilon_s$ is the dielectric constant of the solvent.  This
cavity field then directly leads to the Onsager mean-field
\cite{Hoye:1976} equation for the dielectric constant and, when mutual
short-range correlations of dipoles are included, to the
Onsager-Kirkwood relation \cite{Boettcher:73}. The problem one faces
in an attempt to describe a mixture of nanometer-size solutes with a
molecular solvent is that there is no analog of either of these two
equations. The fundamental line of inquiry here is whether one can
extend equation \eref{eq:17}, or its analog, to such mixtures, or a
new set of rules is required. This question is nontrivial to answer,
even though some initial computer simulations indicate that, indeed,
polarized nanoscale interfaces follow rules different from those
established for cavities carved in dielectrics
\cite{DMepl:08,DMjcp2:11,DMjcp1:12}. The simulations are however
limited by the nanosecond range of time-scales. The question of what
is the polar response of a nanoscale interface at low frequencies
remains therefore open.

This study aims to address this question by analyzing recent
dielectric data obtained for solutions of lysozyme in water
\cite{Cametti:2011ys,Vinh:2011qf}. We first develop a general
formalism that does not anticipate any particular solution for the
local field acting on the protein dipole. As a result of this
analysis, we arrive at a surprising conclusion that the hydration
layers of the protein screen its dipole even more substantially than
anticipated by the standard result for a dielectric cavity given by
equation \eref{eq:17}.

We start with introducing the polarization of the solute-solvent
interface by the combined effect of the external electric field and
the solute dipole moments. This interfacial polarization integrates
into an interface dipole moment, which is assigned to each solute even
in the absence of its own dipole. This development leads to the
equation for the dielectric constant of an ideal solution of
dielectric voids inside the polar liquid. We show that this equation
is quite useful in describing the high-frequency dielectric response
of a real solution, when the relaxation of the solute dipoles is
dynamically frozen. We then proceed to a mixture of polar solutes in a
polar liquid. Here, the cross-correlations of the solute and solvent
dipoles \cite{Boresch:00} are expressed in terms of the cavity-field
susceptibility, which can take different forms depending on the
microscopic structure of the water layer interfacing the solute
\cite{DMjcp2:11}.

\section{Dielectric Response of a Mixture}
Dissolving a polar solute in a polar solvent leads to two distinct
effects on the response of the medium to an external electric
field. The first effect is the exclusion of the solvent from the
volume of the solute. The second effect is the response of the charge
distribution within solute to the orienting torque of the external
field. The two effects are entangled in the polarization of the
interface by the solute charges and by the external field. However,
their contributions to the overall dielectric response of the solution
can be separated in the frequency domain. Since they also originate
from distinct physical interactions, repulsive expulsion on the one hand
and electrostatic interactions on the other, we start with considering
the effect of the solute excluded volume and then add the contribution
of the solute dipole moment to the dielectric response of the
solution.

\subsection{Non-polar Solutes in a Polar Solvent}
Excluding the solvent from the solute volume creates the
solute-solvent interface. From the standard viewpoint of dielectric
theories, any interface carries an interfacial polarization when the
solution is placed in a uniform field of external charges (capacitor
plates of the dielectric experiment). The polarization of the
interface is described in the Maxwell theory of dielectrics by the
surface charge density \cite{Jackson:99}. It is given as the
projection of the dipolar polarization of the dielectric
$\mathbf{P}_S$ at the dividing surface $S$ on the outward normal to
the surface $\mathbf{\hat n}_S$. The surface charge density then
becomes $\sigma_P=\mathbf{\hat n}\cdot\mathbf{P}_S$. This charge
density integrates to a dipole moment of the interface
\begin{equation}
  \label{eq:1}
  \mathbf{M}_0^{\text{int}} = \int_S \mathbf{r}_S
  \sigma_P(\mathbf{r}_S) dS ,
\end{equation}
where the surface integral is taken over the closed surface $S$
enveloping the solute.

The interface dipole polarizes the surrounding dielectric by its own
electric field such that the inhomogeneous Maxwell field
$\mathbf{E}(\mathbf{r})$ around the solute is a sum of the uniform
Maxwell field of the external charges $\epsilon_s^{-1}\mathbf{E}_0$ and the
dipolar field of the polarized interface
\begin{equation}
  \label{eq:2}
   \mathbf{E}(\mathbf{r}) = \epsilon_s^{-1} \mathbf{E}_0 +
  \sum_j \mathbf{T}(\mathbf{r}-\mathbf{r}_j)\cdot \mathbf{M}_0^{\text{int}} .
\end{equation}
Here, $\mathbf{T}(\mathbf{r}-\mathbf{r}_j)$ is the dipolar
tensor describing the electric field at point $\mathbf{r}$ inside the
solvent by a point dipole placed at $\mathbf{r}_j$; the sum runs over
$N_0$ solutes with coordinates $\mathbf{r}_j$.

The Maxwell field $\mathbf{E}(\mathbf{r})$ polarizes the liquid, with
the resulting local inhomogeneous polarization
$\mathbf{P}(\mathbf{r})=(4\pi)^{-1}(\epsilon_s-1)\mathbf{E}(\mathbf{r})$,
decaying to the homogeneous polarization $\mathbf{P}$ of the external
charges far from the solute-solvent interface. The overall dipole
created in the solution is the integral of $\mathbf{P}(\mathbf{r})$
over the volume $\Omega$ occupied by the solvent
\begin{equation}
  \label{eq:3}
   \mathbf{M}_{\text{mix}} =\int_{\Omega} \mathbf{P}(\mathbf{r}) d\mathbf{r}  .
\end{equation}
Here, subscript ``mix'' identifies the solvent-solute
mixture. Assuming that the interfacial dipoles of solutes are independent
of each other, one gets \cite{DMjcp1:12}
\begin{equation}
  \label{eq:4}
  \mathbf{M}_{\text{mix}} = \mathbf{M}_{\text{hom}} - N_0\Omega_0 \mathbf{P} -
  (2/3)(\epsilon_s-1)N_0\mathbf{M}_{0}^{\text{int}} .
\end{equation}
Here, $\mathbf{M}_{\text{hom}}=V\mathbf{P}$ is the dipole moment of the
corresponding homogeneous (without solutes) polarized solvent and
$\Omega_0$ is the volume of the solute;
$\mathbf{P}=(4\pi)^{-1}(1-\epsilon_s^{-1})\mathbf{E}_0$ is the
polarization of the homogeneous solvent. The second summand in equation
\eref{eq:4}  represents the dipole moment cut from the liquid by
inserting $N_0$ voids. Finally, the last term is an additional
polarization induced in the surrounding liquid by the surface charge
density $\sigma_P$.

The value of the interface solute dipole $M_0^{\text{int}}$ will
depend on the specifics of the solute-solvent interactions and the
local polarization of the solvent created by these interactions. While
it is a complex function of the entire mosaic of pairwise
solute-solvent interactions for a realistic solute, an estimate of
this parameter can be obtained from dielectric theories for a
spherical void in a dielectric. The interface dipole reads in this
case \cite{Jackson:99}
\begin{equation}
  \label{eq:5}
   \mathbf{M}_0^{\text{M}} =-3\Omega_0\mathbf{P}/(2\epsilon_s+1) ,
\end{equation}
where the subscript ``M'' specifies Maxwell's electrostatics of a
dividing surface not affected by local solute-solvent interactions. In
order to quantify deviations from this generic result, one can
introduce the ratio 
\begin{equation}
  \label{eq:6}
  \alpha = M_0^{\text{int}}/ M_0^{\text{M}} .
\end{equation}

The dipole moment of the mixture is related to the mixture dielectric
constant $\epsilon_{\text{mix}}$ as
$M_{\text{mix}}/V=(4\pi)^{-1}(1-\epsilon_{\text{mix}}^{-1})E_0$. One
then obtains for the dielectric constant of the mixture
\begin{equation}
  \label{eq:7}
  \frac{\epsilon_s}{\epsilon_{\text{mix}}} = 1 + \eta_0
  (\epsilon_s-1)\left[ 1 - 2\alpha \frac{\epsilon_s-1}{2\epsilon_s +
      1} \right] + R_1(\eta_0) ,
\end{equation}
where $\eta_0=N_0\Omega_0/V$ is the volume fraction of the solutes in
the sample with the overall volume $V$. We have put an extra term
$R_1(\eta_0)$ in the above equation to indicate terms non-linear in
the volume fraction that appear in the dielectric constant when mutual
polarization of the interfacial dipoles is taken into account
\cite{DMpre:10}. Similar non-linear terms appear in the response of a
mixture of water with dipolar solutes discussed below. There is
presently no consistent formalism to include these effects and we
neglect them at the current stage of the theory development
recognizing that the theory might run into conflict with the data
collected for concentrated solutions.

If the Maxwell result for a void in a dielectric holds, $\alpha=1$ and
the dielectric constant of the mixture becomes
\begin{equation}
  \label{eq:8}
   \frac{\epsilon_s}{\epsilon_{\text{mix}}} = 1 + \eta_0
   \frac{3(\epsilon_s-1)}{2\epsilon_s+1} .
\end{equation}
Equation \eref{eq:7}, with $R_1(\eta_0)$ omitted, and \eref{eq:8}
describe the dielectric constant of an ideal mixture of non-polar
solutes and a polar solvent. Equation \eref{eq:8} also reduces to the
standard result of the Maxwell-Wagner theory in the limit of low
volume fraction of the solutes \cite{Takashima:89,Scaife:98}. One can
also account for the electronic polarizability of the protein not
mentioned so far. If the refractive index $n_p$ can be assigned to the
protein, one needs only to realize that the boundary conditions of the
dielectric theories are sensitive to the ratio of the two dielectric
constants at the dividing surface, $\epsilon_s/ n_p^2$. Equation
\eref{eq:8} then extends to
\begin{equation}
  \label{eq:16}
   \frac{\epsilon_s}{\epsilon_{\text{mix}}} = 1 + \eta_0
   \frac{3(\epsilon_s-n_p^2)}{2\epsilon_s+n_p^2} .
\end{equation}

Equation \eref{eq:7} can be alternatively written in terms of the
cavity field $E_c$ inside a spherical void in a uniformly polarized
liquid. The electric field inside the cavity is proportional to the
external field, with the susceptibility $\chi_c=E_c/E_0$. In
terms of this susceptibility, \eref{eq:7} becomes \cite{DMjcp2:11}
\begin{equation}
  \label{eq:21}
  \frac{\epsilon_s}{\epsilon_{\text{mix}}} = 1 + 3 \eta_0
  \left[\chi_c\epsilon_s - 1\right] .
\end{equation}
The standard prescription of Maxwell's  theory of dielectrics predicts
\cite{Boettcher:73,Frohlich} 
\begin{equation}
  \label{eq:22}
  \chi_c^{\text{M}} = \frac{3}{2\epsilon_s+1} .
\end{equation}

The connection between the susceptibility $\chi_c$ and the parameter
$\alpha$ (equation \eref{eq:6}) that is required to obtain equation \eref{eq:21}
from equation \eref{eq:7} is derived from the following arguments. The
polarization $\mathbf{P}(\mathbf{r})$ in the solvent, induced by the
Maxwell field given by equation \eref{eq:2}, creates a non-vanishing
electric field inside the solute that is given by the equation
\begin{equation}
  \label{eq:19}
  \mathbf{E}_c = \mathbf{E}_0 + \int_{\Omega}
  \mathbf{T}(\mathbf{r})\cdot \mathbf{P}(\mathbf{r}) d\mathbf{r} .
\end{equation}
Upon substitution of equation \eref{eq:2} into this relation, one arrives
at the connection between $\chi_c$ and $\alpha$
\begin{equation}
  \label{eq:20}
  3\epsilon_s\chi_c = \epsilon_s + 2 - \alpha \frac{2(\epsilon_s-1)}{2\epsilon_s+1} .
\end{equation}
Combining equations \eref{eq:7} and \eref{eq:20}, one arrives at equation
\eref{eq:21}.

\subsection{Polar Solutes in a Polar Solvent}
\label{sec2-1}
When a solute carries dipole moment $\mathbf{m}_0$, it aligns
along the external field such that the average dipole $\langle m_0
\rangle_E$ in a weak external field is given by linear
susceptibility \cite{Frohlich} $\chi_0$
\begin{equation}
  \label{eq:9}
  \langle m_0 \rangle_E = \chi_0 \Omega_0 E_0 ,
\end{equation}
where $\langle\dots\rangle_E$ denotes an ensemble average in the
presence of the external field and 
\begin{equation}
  \label{eq:10}
  \chi_0 = \chi_{00}+\chi_{0s} = (\beta/3\Omega_0) \langle \delta
  \mathbf{m}_0\cdot \delta \mathbf{M}_{\text{mix}}   \rangle . 
\end{equation}
In this equation, $\delta \mathbf{m}_0 = \mathbf{m}_0 - \langle
\mathbf{m}_0 \rangle $ and $\delta \mathbf{M}_{\text{mix}} =
\mathbf{M}_{\text{mix}} - \langle \mathbf{M}_{\text{mix}} \rangle$ are
the deviations of the solute dipole and the dipole of the sample
$\mathbf{M}_{\text{mix}}$ from their average values and
$\beta=1/(k_{\text{B}}T)$ is the inverse temperature.

The solute susceptibility in equation \eref{eq:10} is split into the self,
$\chi_{00}$, and solute-solvent, $\chi_{0s}$, parts. The former is
given by the variance of a single solute dipole 
\begin{equation}
  \label{eq:11}
  \chi_{00} = (\beta/3\Omega_0) \langle (\delta \mathbf{m}_0)^2\rangle .
\end{equation}
Correspondingly, the cross susceptibility is the correlation of a
single solute dipole with the dipole moment $\delta\mathbf{M}_s$ 
of the entire solvent in the sample \cite{Boresch:00}
\begin{equation}
  \label{eq:12}
  \chi_{0s} = (\beta/3\Omega_0) \langle \delta
  \mathbf{m}_0\cdot\delta\mathbf{M}_s\rangle .
\end{equation}
Equation \eref{eq:11} neglects correlations between dipole moments of
the solutes in the solution represented by the corresponding Kirkwood
factor. Since the latter describes short-range correlations, of the
length-scale of the molecular diameter \cite{SPH:81}, they can be
safely omitted in the type of theory developed here.

Both standard arguments of the dielectric theories \cite{Boettcher:73}
and microscopic derivation \cite{DMjcp1:12} suggest a simple connection
between the solute dipolar susceptibility $\chi_0$ and the self
susceptibility $\chi_{00} $
\begin{equation}
  \label{eq:23}
  \chi_0 = \chi_c  \chi_{00}   .
\end{equation}
This relation implies that the account of the solute-solvent
cross-correlations entering susceptibility $\chi_{0s}$ amounts to
introducing the cavity field acting on the average solute dipoles,
which also defines the torque acting on a selected dipole in the
Onsager theory of dipolar liquids (directing field) \cite{Onsager:36}.

Adding the dipolar polarization of the solutes to equation \eref{eq:21} for
the dielectric constant of the liquid with spherical voids, one
arrives at the dielectric constant of the solution
\begin{equation}
  \label{eq:26}
  \frac{\epsilon_s}{\epsilon_{\text{mix}}} = 1 -3\eta_0 + 
  3\eta_0\epsilon_s\chi_c\left(1 - y_0 \right)  ,
\end{equation}
where $y_0=(4\pi/3)\chi_{00}$.  This equation clearly reduces to
\eref{eq:21} in the limit of non-polar solutes when $y_{0}\rightarrow
0$.

\subsection{Frequency-Dependent Response}
The static arguments presented in the previous sections can be
extended to the frequency domain of main interest to broad-band
dielectric spectroscopy. The dielectric constants of both the solvent
and the mixture become frequency-dependent functions,
$\epsilon_s(\omega)$ and $\epsilon_{\text{mix}}(\omega)$. The dipolar
susceptibility of an isolated solute transforms into a linear response
function, instead of a static correlator of equation \eref{eq:11}. The
relevant formalism is well developed and the result is the following
response function of the solute dipolar
fluctuations \cite{Hansen:03,Nandi:1998qf}
\begin{equation}
  \label{eq:14}
  \chi_{00}(\omega) = \chi_{00}\left[1+ i\omega \tilde
    S_{00}(\omega)\right] .
\end{equation}
Here, $\tilde S_{00}(\omega)$ is the Laplace-Fourier transform of the
normalized time correlation function of the solute dipole
$\mathbf{m}_{0}(t)$
\begin{equation}
  \label{eq:15}
  S_{00}(t)=\left[\langle (\delta \mathbf{m}_0)^2\rangle\right]^{-1} \langle \delta \mathbf{m}_0(t)\cdot
    \delta\mathbf{m}_0(0)\rangle .
\end{equation}
This function was fitted to multi-exponential decay when applied to the
analysis of the MD simulation data presented below
\begin{equation}
  \label{eq:13}
  S_{00}(t) = \sum_i A_i e^{-t/\tau_i},\quad \sum_iA_i =1 ,
\end{equation}
where $\tau_i$ are the relaxation times and $A_i$ are the relative
weights of the relaxation components. From this equation, one gets the 
frequency-dependent function $y_0(\omega)$
\begin{equation}
  \label{eq:144}
  y_0(\omega) = y_0 \sum_i \frac{A_i}{1-i\omega\tau_i} .
\end{equation}
The frequency-dependent dielectric constant of the solution becomes
\begin{equation}
  \label{eq:27}
  \frac{\epsilon_s(\omega)}{\epsilon_{\text{mix}}(\omega)} =  1 -3\eta_0 + 
  3\eta_0\epsilon_s(\omega)\chi_c(\omega)\left(1 - y_0(\omega) \right) .
\end{equation}

Our arguments so far have not included any approximations except
neglecting mutual polarization of solutes at their high concentration
and the short-range correlations of solute dipoles entering the
Kirkwood factor of the solutes. However, equations \eref{eq:26} and
\eref{eq:27} contain an unknown cavity-field susceptibility
$\chi_c(\omega)$. The Maxwell's result for this function refers to a
free surface separating a dielectric from a void. It is a
priory not obvious that this function can describe the complex and
heterogeneous protein-water interface involving both weak
protein-water interactions at hydrophobic patches and strong binding
to charged surface residues. However, one can use the experimental
input for the dielectric constants of the mixture and pure water in equation
\eref{eq:27} to extract the cavity-field susceptibility
$\chi_c(\omega)$.

\begin{figure}
  \centering
  \includegraphics*[width=8cm]{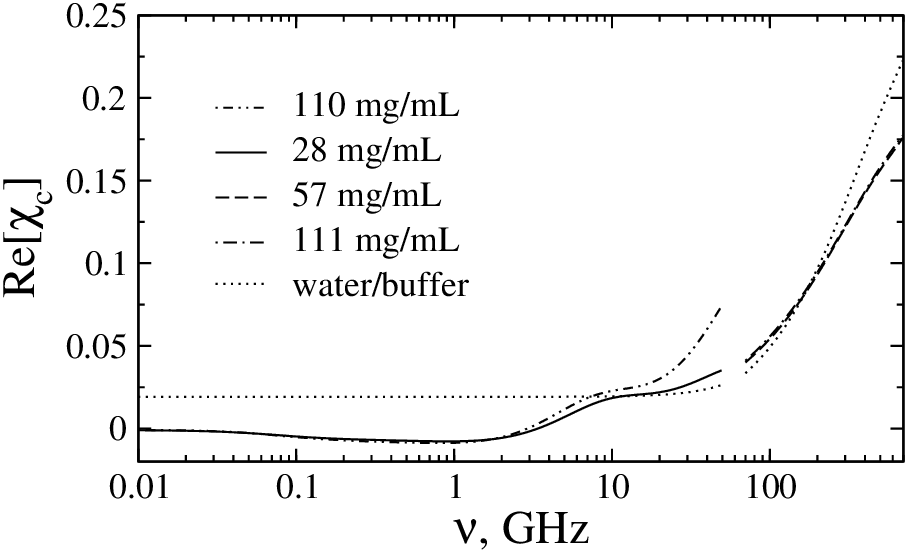}
  \caption{Real part of the cavity-field susceptibility
    $\chi_c(\omega)$ extracted from experimental dielectric
    measurements according to \eref{eq:27}. The results combine the
    broad-band dielectric measurements from \cite{Cametti:2011ys} with
    high-frequency data from \cite{Vinh:2011qf}. The dotted line
    indicates $\chi_c^{\text{M}}$ from equation \eref{eq:22} for pure
    water (at lower frequencies) and the buffer (at higher
    frequencies). The gap between the two sets of curves represents
    the frequency window between the measurements. }
  \label{fig:1}
\end{figure}

Figure \ref{fig:1} shows the real part $\chi'_c(\omega)$ extracted
from equation \eref{eq:27} using frequency-dependent dielectric
constants of lysozyme solutions from broad-band dielectric
spectroscopy below 50 GHz \cite{Cametti:2011ys} and from separate
measurements in the frequency range 70--700 GHz \cite{Vinh:2011qf}. The
dotted line shows $\mathrm{Re}[\chi_c^{\text{M}}]$ from equation
\eref{eq:22}; the break in the curve signals the transition from water
to buffer used at higher frequencies in \cite{Vinh:2011qf}. The
cavity-field susceptibility follows very closely the Maxwell
prediction in the range of frequencies 10--200 GHz, but then deviates
downward outside this range. The behavior at low frequencies is
particularly noteworthy.

It turns out that the dipole moment induced at the protein by an
external field is over-screened \cite{Ballenegger:05} by the hydration
layers, and perhaps the ionic atmosphere, to nearly zero. In fact,
$\chi_c$ is below zero at $\nu < 1$ GHz, implying a die-electric
response, i.e.\ repulsion of the protein dipole from a region of a
stronger electric field. This phenomenon, known as negative
dielectrophoresis, is well-documented for hydrated
nanoparticles \cite{JonesBook:95}, but has not been broadly observed
for proteins. Our recent extensive simulations of
ubiquitin \cite{DMjcp1:12}, which is neutral at pH$=7.0$, have
indicated exactly this scenario: a negative $\chi_{0s}$, larger in
magnitude than the positive $\chi_{00}$, thus resulting in a slightly
negative $\chi_0$ in equation \eref{eq:10}. However, this result has not
been detected by simulations of charged proteins, including lysozyme,
probably due to the neglect of the ionic atmosphere in the analysis.

Figure \ref{fig:1} suggests that dielectric models of the cavity-field
susceptibility do not provide an adequate description in the entire
range of frequencies of interest to broad-band spectroscopy. However,
the modeling can proceed along separate routes since the expulsion of
polar water from the solute core is significant only at high
frequencies, while the polar response of the protein dipole, described
by $y_0(\omega)$, dominates at low frequencies. One therefore can keep
the Maxwell result for $\chi_c(\omega)$ for the former component, as
realized in equation \eref{eq:8}. Since there is currently no model
allowing to describe the overscreening observed at low frequencies, we
have resorted to an empirical approximation.  Replacing
$\chi_c\epsilon_s y_0$ in eqs \eref{eq:26} and \eref{eq:27} with
$\chi_c^{\text{M}} y_0$ accomplishes most of what is seen to occur in
figure \ref{fig:1} and allows us to arrive at a compact relation for the
dielectric constant of the solution
\begin{equation}
  \label{eq:18}
  \frac{\epsilon_s(\omega)}{\epsilon_{\text{mix}}(\omega)} = 1 +
  \frac{3\eta_0}{2\epsilon_s(\omega)+1}\left[\epsilon_s(\omega) - 1 -
    3 y_0(\omega)  \right] .
\end{equation}

\subsection{Dielectric instability}
Equation \eref{eq:26} predicts a point of dielectric instability at
which the assumption of a uniform solution of weakly interacting
protein dipoles breaks down. The instability is toward clustering of
dipoles and is associated with the divergence of the dielectric
constant $\epsilon_{\text{mix}}$. It is reached at the critical volume
fraction 
\begin{equation}
  \label{eq:28}
  3\eta_c = \left[1 +\epsilon_s\chi_c(y_0-1)\right]^{-1} .
\end{equation}
If the Maxwell form of the cavity-field susceptibility is used in this
equation, the critical point $\eta_c = 0.01$ ($y_0\simeq 16$)
corresponds to the concentration of 8 mg/mL for lysozyme in
solution. Lysozyme solutions are stable in this range of
concentrations and this estimate is clearly too low. On the contrary,
the overscreening scenario shown in figure \ref{fig:1} makes $\eta_c$
negative, thus removing the instability altogether. While other forms
of aggregation are still possible \cite{Liu:2010vn,Cardinaux:2011uq},
it might be quite possible that overscreening of the protein dipole
eliminates instability toward dipolar clustering (such as formation of
dipolar chains) and lowers the sensitivity of proteins in solutions to
inhomogeneous electric fields always present \textit{in vivo}.

\begin{figure}
  \centering
  \includegraphics*[width=8cm]{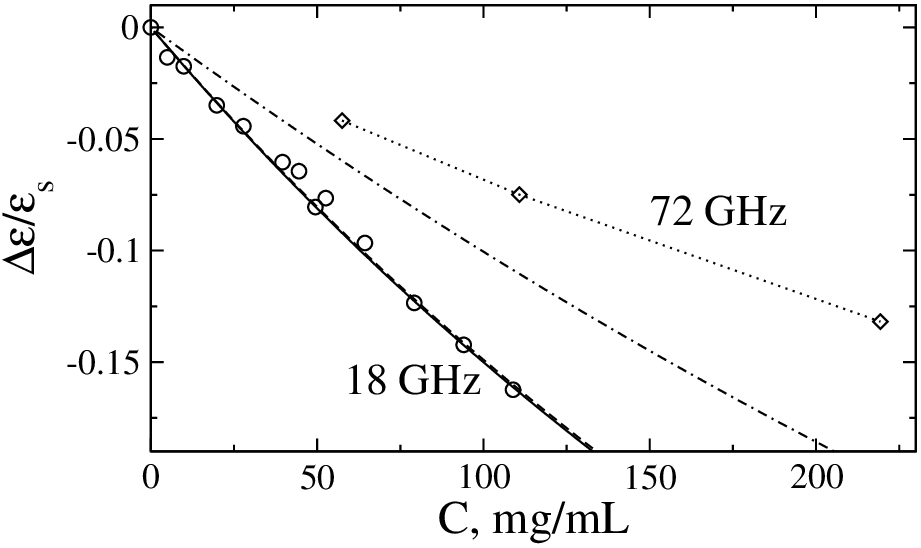}
  \caption{Decrement of the water dielectric constant in the solution
    of lysozyme in water, $\Delta \epsilon(\omega)/ \epsilon_s(\omega)
    = \epsilon_{\text{mix}}(\omega)/\epsilon_s(\omega) -1$, as a
    function of the protein concentration $C$. The points are the
    experimental data at the frequency of the water Debye peak
    $\nu_{\text{D}} \simeq 18$ GHz (circles) \cite{Cametti:2011ys},
    and at $\nu = 72$ GHz (diamonds) \cite{Vinh:2011qf}. The solid and
    dash-dotted lines refer to the calculations using equations \eref{eq:8}
    and \eref{eq:16} in the order of increasing frequency.  The dashed
    line is the calculation incorporating the dynamics of the protein
    dipole according to equation \eref{eq:18} with $y_0(\omega_D)$
    calculated from MD simulations \cite{DMjcp1:12}. The lysozyme
    molecular volume of $\Omega_0=29.8$ nm$^3$ is used to convert from
    the volume fraction to the solution concentration. The dotted line
    connects the experimental points. }
  \label{fig:2}
\end{figure}

\section{Application to Experiment: Lysozyme Solution}
Dielectric measurements of solutions typically provide the real and
imaginary parts of the dielectric constant as functions of frequency
and solution composition
\cite{Oleinikova:04,weingartner:1463,Cametti:2011ys,Tielrooij:10,Rahman:2011fk}.
The existence of these two coordinates, frequency and solute
concentration, allows one to learn about the specific pattern of
interfacial polarization realized for a given solute and the dynamics
of processes contributing to the relaxation of the sample dipole
moment. We start with the analysis of the concentration dependence at
a given frequency, followed with the analysis of the frequency
dependence at a fixed concentration.

\subsection{Decrement of the water Debye peak}
Independently of the details of the dynamics of a protein itself and
its coupling to the interfacial waters, the time-scales of these
motions are significantly lower than the characteristic time of
dielectric relaxation of water. The global motions of the solute are
dynamically frozen at the frequency of the water Debye peak
($\nu_D\sim 18$ GHz).  This implies that $y_0(\omega_D)$ can be
dropped from equation \eref{eq:18}. One then arrives at the dielectric
constant of the mixture of polar water and effectively non-polar
solutes (eqs \eref{eq:7} and \eref{eq:8}). Any sufficiently high
frequency can in principle be taken for this analysis. The decrement
of the Debye peak of water in the solution vs the solute concentration
is often reported \cite{South:72} and can be used, in the framework of the present
theory, as a convenient source of data to extract the information
about the parameters $\alpha$ and $\chi_c$.

Our formalism is applied to recent measurements of dielectric spectra
of lysozyme solutions \cite{Cametti:2011ys,Vinh:2011qf}. Figure
\ref{fig:2} shows the dependence of the decrement in the amplitude of
the water Debye peak $\omega_{\text{D}}$ in the solution $\Delta
\epsilon(\omega_{\text{D}}) =\epsilon_{\text{mix}}(\omega_{\text{D}}) -
\epsilon_s(\omega_{\text{D}})$ vs the protein concentration. Circles
show the experimental data from \cite{Cametti:2011ys}, while the solid
and dashed lines refer to equations \eref{eq:8} and \eref{eq:18},
respectively. For the latter, $y_0(\omega_{\text{D}})$ calculated from
MD simulations \cite{DMjcp1:12}, and discussed below for the analysis
at lower frequencies, was used. Clearly, the protein permanent dipole can
be safely neglected. The transformation from the solution
concentration reported experimentally to the volume fraction required
by \eref{eq:8} and \eref{eq:27} was performed by using the volume
of lysozyme $\Omega_0=29.8$ nm$^3$. The latter was calculated from the
crystallographic structure of the protein (3FE0, PBD database) by
using the algorithm developed by Till and Ullmann \cite{Till:10}.

In accord with the results shown in figure \ref{fig:1}, the cavity-field
susceptibility is well described by the Maxwell form (equation
\eref{eq:22}) at the frequency of the water Debye peak, and the
agreement between theory and experiment is excellent. It becomes less
satisfactory at a higher frequency of 72 GHz \cite{Vinh:2011qf}, also
shown in figure \ref{fig:2}. The refractive index of the protein
starts to affect the result at this high frequency and $n_p=1.7$ from
\cite{Knab:06} was adopted in the calculations using equation
\eref{eq:16}. The theoretical slope with increasing protein
concentration is higher than experimentally reported and is likely
related to deviations from the Maxwell form of the cavity-field
susceptibility seen in figure \ref{fig:1}.

\begin{figure}
  \centering
  \includegraphics*[width=8cm]{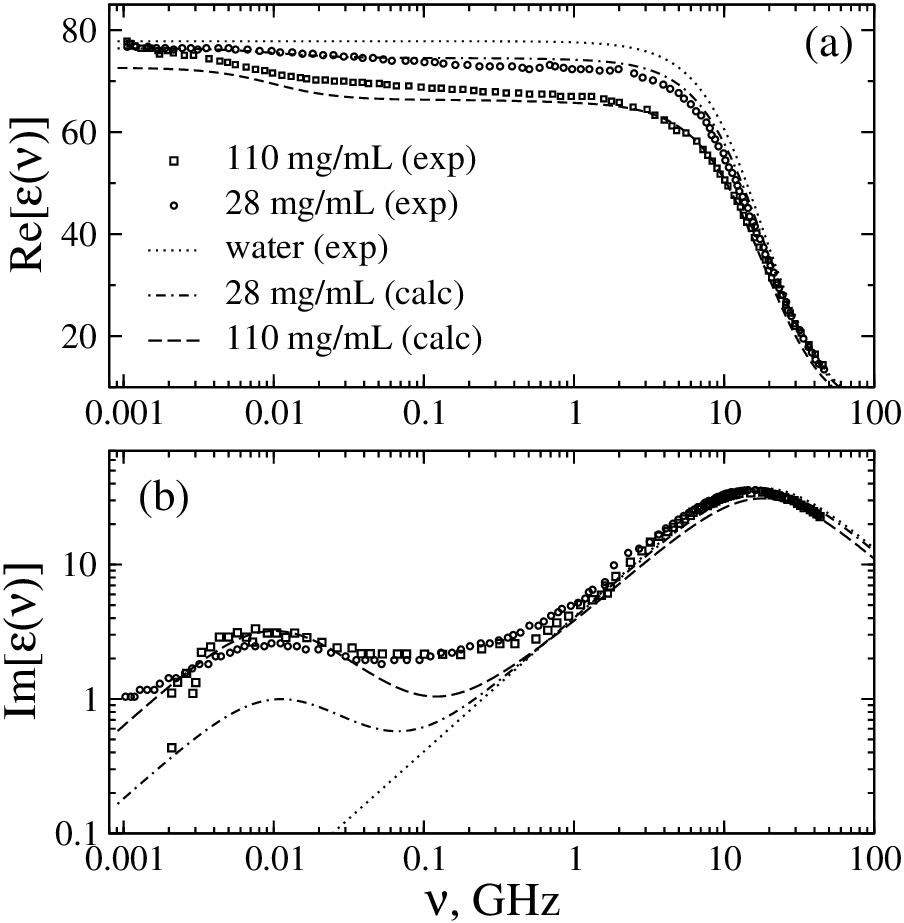}
  \caption{Real (a) and imaginary (b) parts of the dielectric constant
    of the lysozyme solution measured experimentally
    \cite{Cametti:2011ys} (points) and calculated theoretically
    (lines). The dotted lines show the real and imaginary parts of the
    dielectric spectrum of water from \cite{Yada:09}. }
  \label{fig:3}
\end{figure}

\subsection{Dielectric spectra of solutions}

The results for the dispersion and loss spectra of lysozyme solutions
are shown in figure \ref{fig:3}.  Experimental data from
\cite{Yada:09} were used for $\epsilon_s(\omega)$ and Molecular
Dynamics (MD) simulations of a single lysozyme protein hydrated in a
simulation box of TIP3P waters \cite{DMjcp1:12} were used to produce
$y_0(\omega)$ in \eref{eq:144}. The relaxation parameters in
\eref{eq:144} are: $A_i=\{0.13, 0.06, 0.81 \}$, $\tau_i=\{0.037,0.295,
14.6\}$ ns, $y_0=16.3$. The dominant relaxation component of the
solute dipole, with the relaxation time of $14.6$ ns, can be assigned
to protein tumbling. The relaxation time of 9.1 ns was reported for
this relaxation component from the analysis of proton NMR at low
resonance frequencies \cite{Krushelnitsky:06}.

The usefulness of MD simulations is somewhat limited for the sake of
comparison with experiment since the charge distribution in the
protein studied by simulations might not entirely fit the experimental
conditions. The standard force-field prescriptions for
protonating/deprotonating the surface residues of lysozyme at
$\mathrm{pH}=7.0$ produce the overall protein charge of $+7$, while
the charge of $+10$ is reported at pH$=5.5$ in the experimental study
\cite{Cametti:2011ys}. Overall, permanent dipole moments of proteins
arise from slight deviations from highly symmetric distribution of
charge minimizing the total dipole moment
\cite{Barlow:87,Takashima:02}. Shifts of p$K_a$ values of surface
residues due to local electrostatic environment \cite{Mellor:11}, ion
association, and pH can therefore alter the dipole moment.

Despite remaining uncertainties regarding the magnitude of the protein
dipole when experimental conditions are concerned, the dipole $\langle
m_0 \rangle=223$ D from MD results in a fair agreement between
theoretical and experimental dispersion curves
$\epsilon'_{\text{mix}}(\omega)$ at the lower concentration of the
protein, 28 mg/mL (figure \ref{fig:3}a). The theory misses some of the
static dielectric constant at the higher concentration, $c=110$ mg/mL,
but the difference actually comes from the missing increment at
intermediate frequencies associated with $\delta$-dispersion. This
part of the spectrum is also missing in the loss spectrum (figure
\ref{fig:3}b).  This outcome is expected since specific protein-water
binding contributing to this signal
\cite{Boresch:00,Weingartner:01,Oleinikova:04} has not been
incorporated into the model.

\section{Summary}
Broad-band dielectric spectroscopy is a widely used tool to
interrogate the dynamics of complex systems, including protein
solutions. The interest in the field in the recent years has been to
extract the polarization properties and dynamics of protein hydration
layers from frequency-dependent spectra. The standard approach to the
problem is to fit the dispersion and loss spectra to a sum of Debye or
stretch-exponential functions, assuming that each component represents
a separate relaxation process in a complex environment.  The obvious
limitation of this approach is the non-additivity of interfacial
polarization, well recognized by classical theories of dielectric
mixtures \cite{Takashima:89,Scaife:98,Choi:99}. While these classical
theories were developed for mixtures of dielectric materials, when
each component can be represented by a macroscopic dielectric body,
their application to hydrated proteins is clearly limited. At the same
time, standard theories of dipolar liquids \cite{Madden:84,SPH:81} are
not of much use either since they do not recognize the existence of an
extended polarizable interface, which is in fact the central concept
of the dielectric theories of mixtures. The present theoretical
development aims to fill the void existing in each approach by
recognizing both the molecular nature of the protein dipole and a
quasi-macroscopic subensemble of interfacial waters producing
interfacial polarization.

The theory thus aims to study if the standard rules established for
cavities carved in dielectrics, and also applied to calculate the
local field acting on molecular dipoles \cite{Onsager:36}, can be
applied to hydrated proteins. Equations \eref{eq:27} is central to
this analysis since it allows us to extract the cavity-field
susceptibility $\chi_c(\omega)$ directly from the frequency-dependent
dielectric constants of the protein solution and pure water.  The
remarkable result of this analysis is that at $\omega < 1$ GHz the
susceptibility $\chi_c(\omega)$ is below $\simeq 0.02$ predicted by
the Maxwell equation \eref{eq:22} and is in the negative territory,
down to $\simeq -10^{-3}$. Therefore, the standard prescription
derived for dielectric cavities (equations \eref{eq:17} and
\eref{eq:22}) cannot be used in successful theories of dielectric
response of protein solutions.

Granted, the cavity susceptibility extracted from experimental
measurements might reflect the combined response of the dielectric
interface and the ionic atmosphere. However, as a cumulative signature
of the protein-water interface, it dramatically downscales the
permanent dipole sensed by the dielectric experiment compared to its
value calculated from atomic charges. Its low value can also help to
explain the puzzling ability of proteins to stay in solution
\textit{in vivo}, despite significant electric field gradients that
should pull a paraelectric particle to stick to, for instance, the
bilipid membrane. The die-electric response suggested by the present
analysis of experimental data, and our previous simulations
\cite{DMjcp1:12}, might be an answer to this puzzle since a
die-electric solute repels from a charged interface creating the field
gradient. It also eliminates the dielectric instability toward
clustering of the solute dipoles predicted by \eref{eq:26} and 
\eref{eq:28} when the Maxwell form of the cavity-field susceptibility
is used there.

\ack This research was supported by the National Science Foundation
(DVM, CHE-0910905). CPU time was provided by the National Science
Foundation through TeraGrid resources (TG-MCB080116N). The author is
grateful to Drs.\ Cametti and Nguyen for sharing their experimental
results.

\section*{References}

\bibliographystyle{unsrt}
\bibliography{chem_abbr,dielectric,dm,statmech,protein,liquids,solvation,dynamics,glass,elastic,simulations,surface,bioet,et,nano,photosynthNew,enm,bioenergy,nih}

\end{document}